\documentclass[11pt]{article}
\usepackage{graphicx}
\usepackage{hyperref}
 
\begin{document}

\begin{center}
\textbf{\LARGE Author Identifiers in Scholarly Repositories}\vspace*{3ex}\\
Simeon Warner\\
Cornell Information Science and\\
Cornell University Library\\
Ithaca, NY 14850, USA\\
\textsf{simeon.warner@cornell.edu}\vspace*{0.5ex}\\
{\small Submitted: 2009-10-09}
\end{center}

\section*{Abstract}

Bibliometric and usage-based analyses and tools highlight the value of 
information about scholarship contained within the network of authors, 
articles and usage data. Less progress has been made on populating and 
using the author side of this network than the article side, in part 
because of the difficulty of unambiguously identifying authors. I 
briefly review a sample of author identifier schemes, and 
consider use in scholarly repositories. I then describe preliminary work 
at arXiv to implement public author identifiers, services based on them, 
and plans to make this information useful beyond the boundaries of arXiv.

\section{Context}

In an ideal scholarly communication system there would be tools 
to browse, navigate, make recommendations and assess influence based on 
the complete graph of all actors (people, collaborations, institutions) 
and all communication artifacts (articles, comments, blog posts, usage 
data\footnote{Logically usage data would be links between actors and 
artifacts. However, for historical, cultural and practical reasons most
usage data is treated as anonymous even though co-usage information may be extracted.}). 
As a shorthand I will call this complete graph the \textit{publication network}.
Contained within it are the familiar citation, usage, co-authorship, and 
co-citation graphs. In recent bibliometric and usage-based work, significant 
progress has been made with the artifact part of this graph (see, for 
example the work of the MESUR project~\cite{BOLLEN+08}). Much less 
progress has been made with the actor part of the graph, in part 
because it is much harder to unambiguously identify authors than articles.

\begin{table}[ht]
\small
\begin{center}
\begin{tabular}{ll}
\hline
Lastname, Initial & Count \\
\hline
Zhang, Y & 100 \\
Lee, J   & 97  \\
Wang, Y  & 89  \\
Wang, J  & 84  \\
Chen, Y  & 77  \\
Kim, J   & 77  \\
Wang, X  & 76  \\
Lee, S   & 74  \\
Kim, S   & 69  \\
Liu, Y   & 69  \\
\hline
\end{tabular}
\end{center}
\caption{\label{tab-names} Most frequently occurring ${lastname, initial}$ 
pairs in arXiv user accounts. There may be a few duplicate accounts but this 
indicates that nearly 100 \textit{different} people named ``Zhang, Y'' have 
created user accounts at arXiv (as of May 2009).}
\end{table}

Consider table~\ref{tab-names} which shows the most frequently occurring
${lastname, initial}$ pairs in arXiv user accounts. This illustrates one facet of
the name disambiguation problem, namely that there are many authors with the 
same name. This is compounded by inconsistent spellings, use of initials or 
full first names, and even name changes. Within a single repository such
as arXiv it is not usually possible to accurately answer the question 
``show me all the articles by \textit{this} Zhang, Y''. In recent years there has 
been considerable work on unsupervised and supervised author name disambiguation 
using many different heuristic, machine learning and clustering techniques,
and many different properties including co-authorship, citations and subjects/topics.
While much better than naive approaches, these techniques are still far from 
perfect.

In a recent Nature Correspondence, Raf Aerts asked
\textit{``If it is possible to have DOIs for objects (or, so they say, 
enough IPv6 addresses for every molecule on Earth), why is it so difficult 
to implement DAIs [Digital Author Identifiers] for authors?''}~\cite{AERTS08}.
Raf had earlier hinted at part of the answer by pointing out that he 
has more than one identifier in Scopus~\cite{ScopusAuthorId}. As we have already
discussed, it is difficult to mine existing data to disambiguate references 
to authors. The more fundamental part of the answer is that it is much 
easier to create DOIs for articles when the one owner for an article 
creates the one DOI for it and presents it with 
the article (ignoring the issue of multiple versions of articles). 
As authors, we are not owned by a single authority and even if an identifier
were created for us at birth by the appropriate government, there would be 
significant privacy concerns about using it for everything. Consider, for 
example, concerns over the uses and misuses of social security 
numbers in the USA. While we want to link a single author's works together, 
do we want that identity to immediately link us to all other digital information 
about the private life of the individual?
 
\section{Author Identifiers}

To illustrate the diversity of currently used author identifiers, 
table~\ref{tab-id-schemes} shows several example schemes used in the
scholarly domain. A more detailed inventory is provided on the \textsf{repinf} 
wiki~\cite{REPINF_Author-identification_2009-10-09}. 
The OpenID and ISNI schemes are not limited to the scholarly domain. OpenID
is aimed primarily at authentication, however, if it continues to see growing 
acceptance it may well be a useful open system that repositories could use. 
It is not clear whether ISNI will develop into a widely used system.
The largest efforts to create author identifiers specifically for the scholarly 
domain, Scopus Author Identifiers and ResearcherID, come from commercial entities 
and are clearly motivated by the desire to provide improved services based upon 
then. It is not clear how open the interfaces based on these identifiers will be, 
or what data about them will be openly available.

\begin{table}[ht]
\small
\begin{tabular}{p{1.2in}p{2.2in}p{0.8in}p{1.6in}}
\hline
Scheme & Example & Scope & Authority \\
\hline
\hline
\href{http://openid.net}{OpenID} &
 \url{http://samruby.myopenid.com/} &
 People &
 Distributed, anyone supporting the protocol, relies upon DNS \\ 
\hline
\raggedright\href{http://www.isni.org/}{ISNI} (International Standard Name Identifier) &
 \texttt{ISNI 1422 4586 3573 0476} &
 People & 
 Draft ISO standard requiring central DB operated by proposed International Agency \\
\hline
Scopus Author Id &
 \texttt{7103063073} & 
 Academia &
 Elsevier \\
\hline
\href{http://www.researcherid.com}{ResearcherID} &
 \texttt{A-1637-2009} & 
 Academia &
 Thomson Reuters \\
\hline
Digital Author Id &
 \texttt{info:eu-repo/dai/nl/304825271} &
 Dutch \mbox{Researchers} &
 Dutch Universities and Research Institutes \\
\hline
\raggedright RePEc Author \mbox{Service} &
 \texttt{pzi1} &
 Economics &
 RePEc \\
\hline
arXiv Author Id &
 \url{http://arxiv.org/a/warner\_s\_1} &
 arXiv.org &
 arXiv.org \\
\hline
\end{tabular}
\caption{\label{tab-id-schemes}A sample of identifier schemes used for scholarly author identifiers} 
\end{table}

The three other examples in table~\ref{tab-id-schemes} illustrate decreasing scopes. 
In the Netherlands the Digital Author Id (DAI) \textit{``is a unique national number 
assigned to every author who has been appointed to a position at a Dutch university 
or research institute or has some other relevant connection with one of these 
organizations''}~\cite{DutchDAI}. The DAI provides a join point for data in 
different repositories and enables services based on this combined data (e.g. 
\href{http://www.narcis.info/person/query/meeus/RecordID/PRS1237369/Language/nl}{NARCIS}).
The RePEc scheme identifies authors and is used to link their publications 
together within the RePEc system serving the economic community. The 
AuthorClaim project aims to extend the RePEc model, using the same software 
infrastructure, to the entire academic domain. Finally, the arXiv author 
identifier, described below, is local to a single repository.

\begin{figure}
\begin{center}
\includegraphics[width=4in]{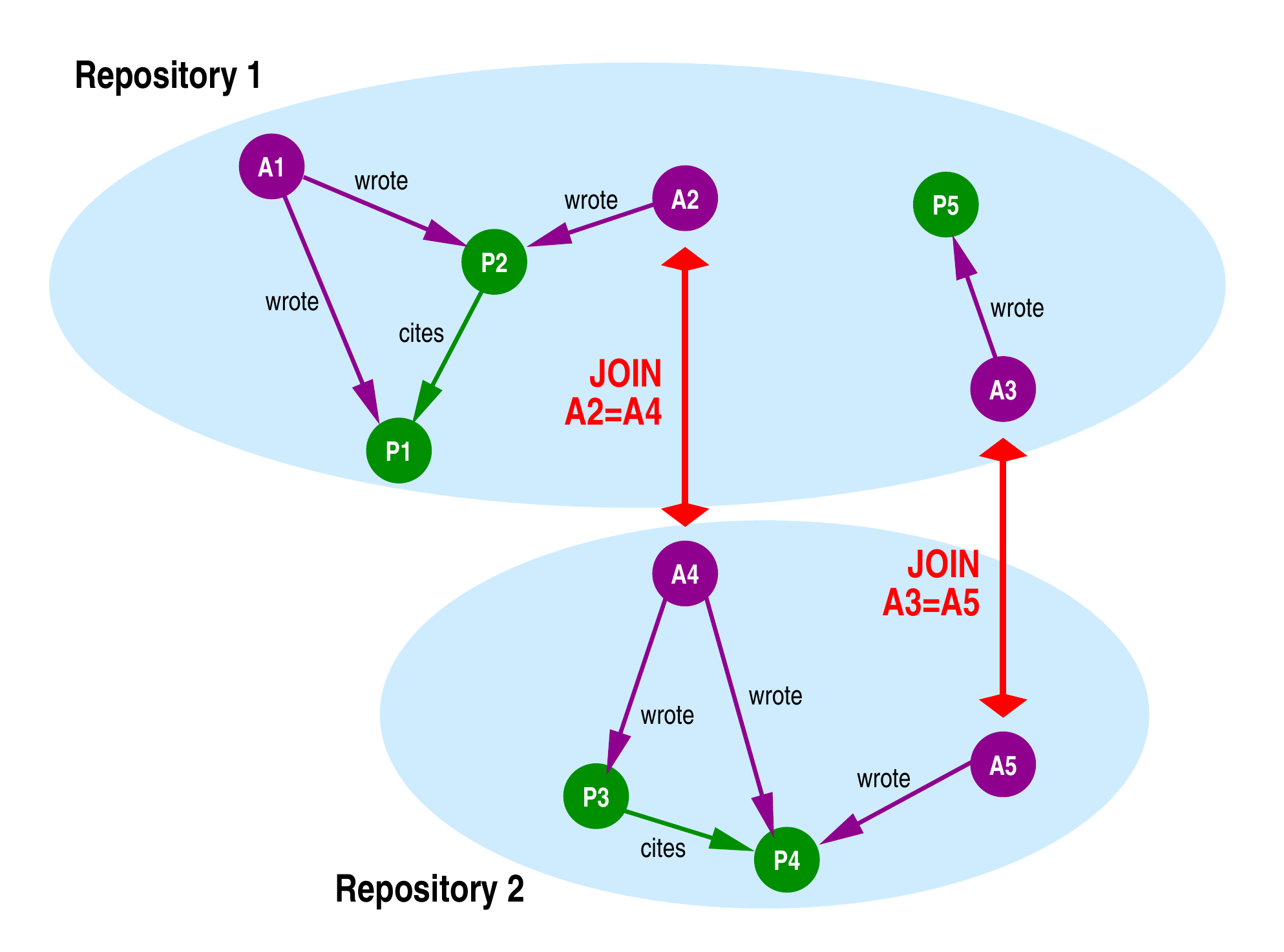}
\caption{\label{fig-authors-papers}If authors A2 and A3 in repository 1 are identified 
with authors A4 and A5 in repository 2 respectively, then the components of the 
publication graph from the two repositories can be joined. The P\# nodes indicate papers.}
\end{center}
\end{figure}

The arguments above and the understanding that there are many different 
interests in, and uses for, author/person identities, suggest that there will 
be many different systems and multiple identities for each author. In the 
scholarly communication domain there will be a patchwork of overlapping 
publication networks. Unless one system grows to dominate, the different 
patches of publication network will identify authors using different 
identifiers. However, it will be vastly easier to match multiple identifiers
for each author than to disambiguate multiple authors with the same name.
A significant aid will be the addition of assertions that link identities 
in different networks (e.g. \textsf{Author A2} in \textsf{Repository 1} 
is the same person as \textsf{Author A4} in \textsf{Repository 2}) as 
illustrated in figure~\ref{fig-authors-papers}. This linking information might
be expressed either via the Semantic Web or in repository metadata.
The ability to record foreign identifiers in the author record within a 
repository, and the ability to match articles between repositories, will 
allow the joining of data across repository boundaries.

\section{Author Identifiers at arXiv}

There are a significant number of physicists for whom all articles, or 
at least all recent articles, are available on arXiv. It is not uncommon 
to find web homepages with a link to arXiv author search in place of 
a bibliography --- why maintain the information in a second place 
when arXiv will do it automatically? Fielded author search has been used 
in this way for many years and has exactly the same problems of author 
disambiguation as text-based efforts to build the publication network.

With the introduction of user accounts, arXiv, like many other repositories,
started to collect data on which user made each submission and whether he or she
claimed to be an author. This start to building authority records was augmented by
attempts to retrospectively associate older papers with users based on email 
address matching, and the introduction of facilities by which users could
``claim ownership'' of existing submissions. Use of the claim ownership facility
was motivated through the introduction of an endorsement 
system\footnote{\url{http://arxiv.org/help/endorsement}} where users must be 
known as authors of a certain number of papers in order to endorse new users.
Various heuristics are used to limit what papers can be claimed automatically
and so far these have proved adequate to avoid incorrect claims being automatically
accepted. 

Demanding identification of all authors at submission time was considered 
impractical. For articles with one or two authors identification would not 
be too burdensome, but for papers with 10 or even 2500 
authors\footnote{Articles from high-energy physics collaborations often have
many authors. See, for example, the recent ATLAS collaboration paper 
\url{http://arxiv.org/abs/0901.0512} with $>2500$ authors} it is clearly 
impractical. A solution that uses arXiv administrator effort to deal with 
each article is also impractical because just two administrators handle all user 
queries relating to arXiv's 58,000 submissions/year --- most submissions must 
be entirely automated. We thus decided on an approach that will create useful 
services based on a public author identifier which we internally link to our 
user records. We hope that by providing useful services our users will be 
motivated to further improve the authority records on which these services 
depend.

\subsection{Author URI and Services}

\begin{figure}
\begin{center}
\includegraphics[width=4in]{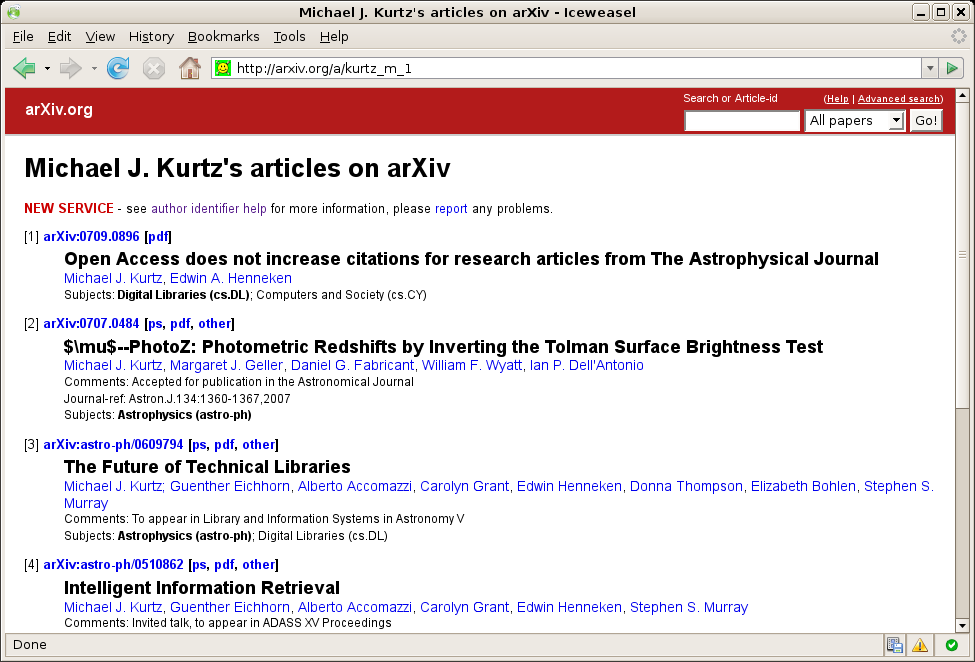}
\caption{\label{fig-id-html}HTML screen returned when an arXiv author id is accessed and 
HTTP content negotiation results in HTML, or HTML is explicitly requested
by appending \texttt{.html} to the author id (\url{http://arxiv.org/a/kurtz_m_1.html})}
\end{center}
\end{figure}

\begin{figure}
\begin{center}
\includegraphics[width=4in]{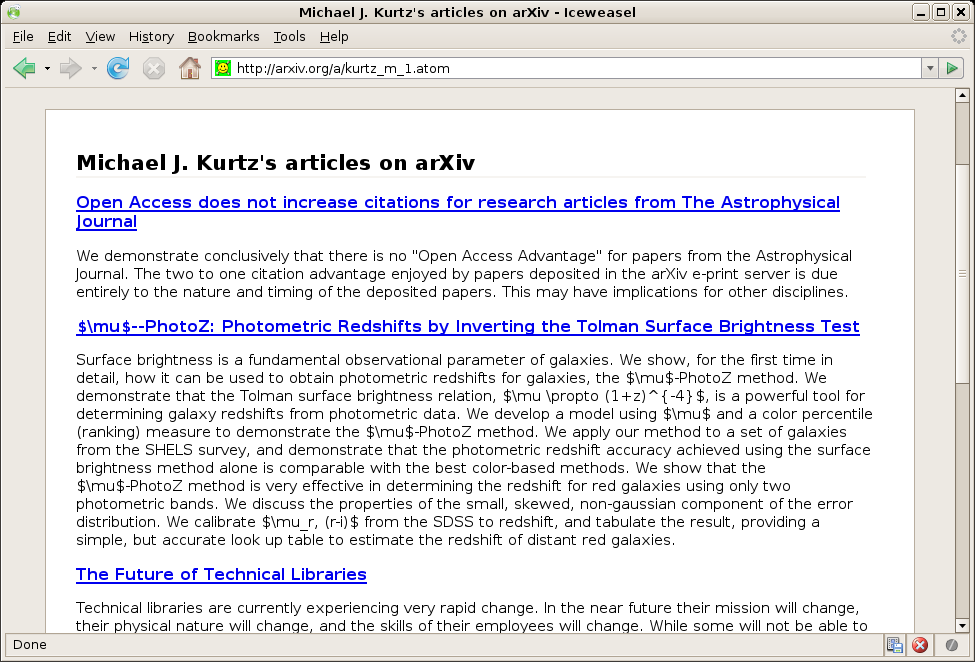}
\caption{\label{fig-id-atom}Web browser (Firefox) rendering of the Atom feed 
returned when an arXiv author id is accessed and HTTP content negotiation 
results in Atom, or Atom is explicitly requested by appending 
\texttt{.atom} to the author id (\url{http://arxiv.org/a/kurtz_m_1.atom})}
\end{center}
\end{figure}

We have opted for a web-centric approach using Linked Data~\cite{BIZER+07} style HTTP
access. Each arXiv author identifier is a unique URI 
(e.g. {\small \url{http://arxiv.org/a/warner\_s\_1}})
which supports HTTP content-negotiation. These URIs are designed to be
human copyable and are based on an ASCII dumb-down of the author name.
By default, or if selected via content-negotiation headers, the arXiv 
author URI redirects to an HTML page listing all arXiv publications 
authored by the given individual based on our user records. An example 
HTML page is shown in figure~\ref{fig-id-html}.
In cases where the author-article associations are complete this facility 
already  solves the problem of name collision in arXiv author search and 
so provides a more reliable link than our text-based author search.
To allow the data to be used by other applications or to allow display 
or monitoring with a feed reader, the list of articles associated with 
an author id is also available as an Atom feed. Figure~\ref{fig-id-atom} 
shows that same data as figure~\ref{fig-id-html} but rendered from the Atom 
feed. As of September 2009 we see about 300 different author id 
URIs being accessed per week to return HTML pages or Atom feeds.

\begin{figure}
\begin{center}
\includegraphics[width=3in]{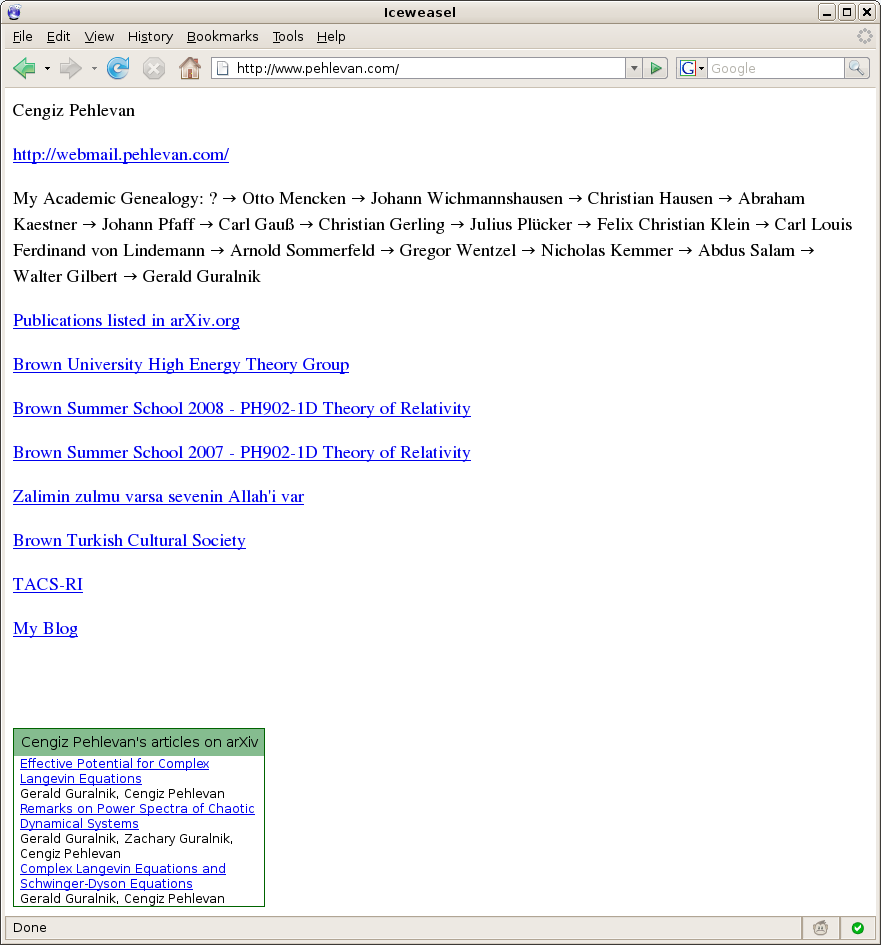}
\includegraphics[width=3in]{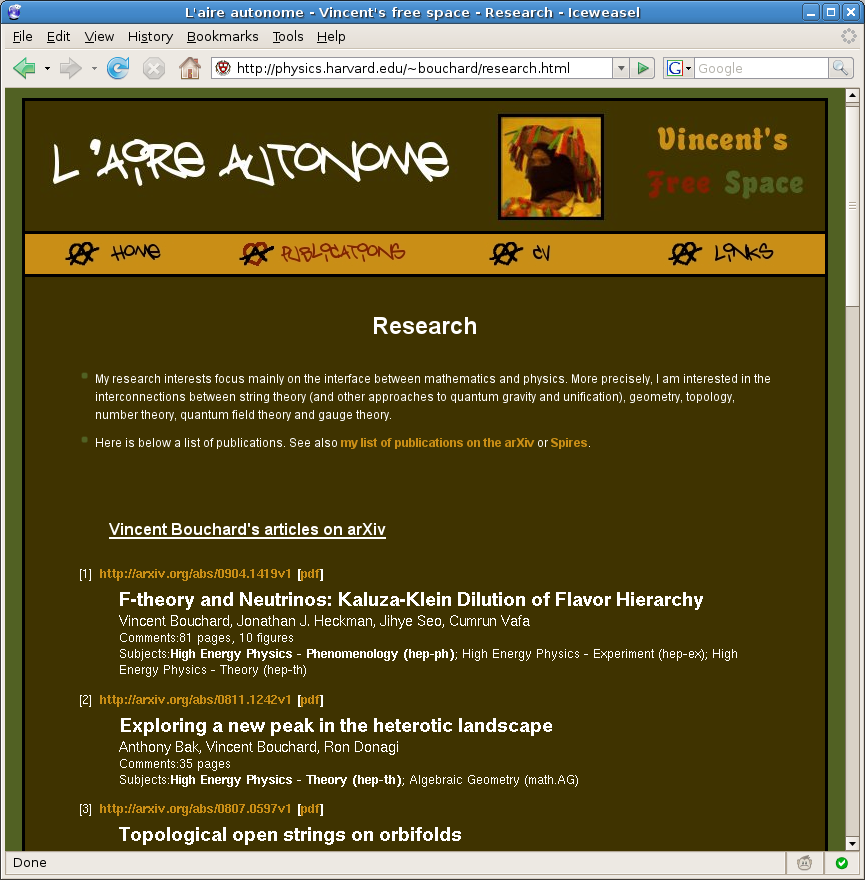}
\caption{\label{fig-myarticles}Two early adopter examples of use of 
the \texttt{myarticles} widget. The example on the left shows the 
``Google ads'' formatting option used to produce a compact display 
in the lower left corner of the browser window. The example on
the right shows the ``arXiv list'' formatting option which picks
up local stylesheet information. In both cases the data from arXiv is 
seamlessly embedded in the user's homepage.}
\end{center}
\end{figure}

A list of articles on the arXiv site is still one click away from the user's 
homepage. We thus provide JavaScript code, which we call the \texttt{myarticles} 
widget, that a user may include in their personal homepage to dynamically include 
an up-to-date publication list from arXiv. Various formatting
options are provided and the content may be styled using CSS. 
Figure~\ref{fig-myarticles} shows two screen shots from early adopters 
of the \texttt{myarticles} widget.
This facility is based upon a content-negotiated request for an Atom 
representation of the arXiv author id resource which results in a machine 
readable Atom feed of paper information (in the same format as the arXiv 
API\footnote{\url{http://arxiv.org/api}}).

\begin{figure}
\begin{center}
\includegraphics[width=4in]{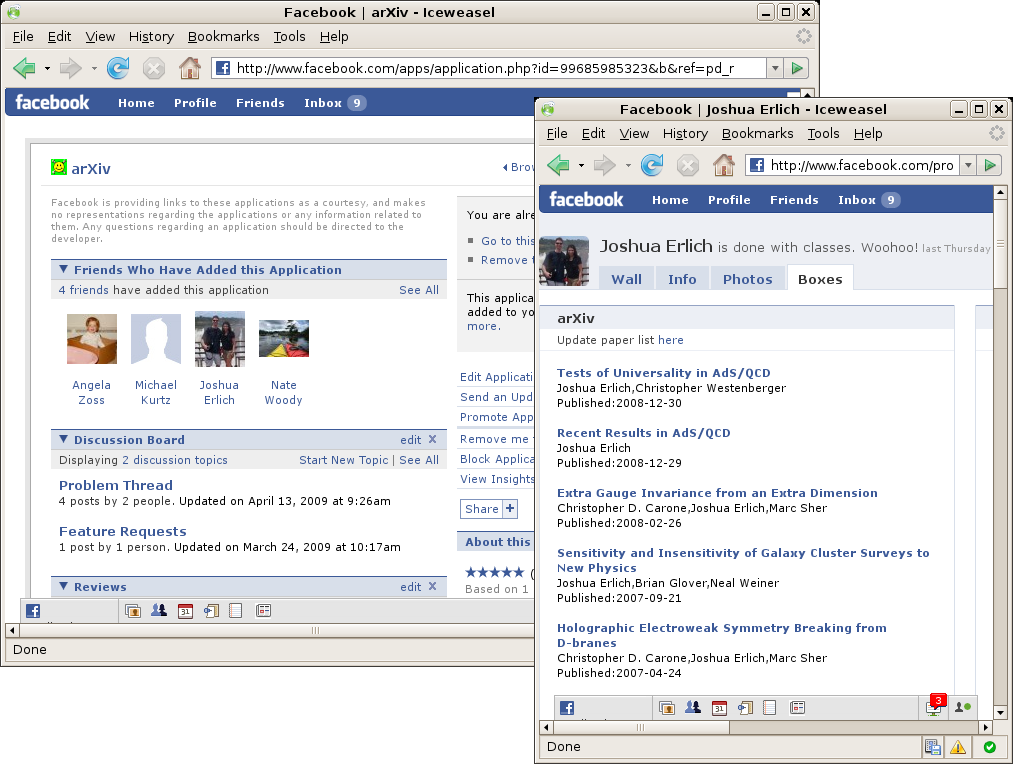}
\caption{\label{fig-fb}Example screenshots from the arXiv Facebook application. 
The window on the right shows a list of article titles and authors that is 
automatically generated because Joshua Erlich enabled the arXiv application 
from with his Facebook account and told it the association with his arXiv 
author id. Any new arXiv article owned by him will automatically be
added to this list.}
\end{center}
\end{figure}

arXiv's second use of arXiv author ids is to leverage this automatically 
generated and updated list of publications to lower the effort required to 
integrate arXiv papers into social networking sites. Facebook was chosen
as the first site to work with but the OpenSocial API is also being 
investigated. Once the arXiv Facebook application has been told the association 
between a user's Facebook account and their arXiv author identifier, a list of 
publications is immediately available as either a panel or a tab on their 
Facebook profile as shown in figure~\ref{fig-fb}. 
All title, author list, abstract and linking information
is automatically imported from arXiv. New or old publications may be reported
in the user's feed, with optional comments, and thus show up in friends' news
feeds. This application was released in March 2009 and sees approximately 600 users
per week as of September 2009. Use is steady and increasing, but there
is not rapid adoption. We continue to experiment with new facilities and
modes of interaction.

\subsection{Helping to Build the Publication Network}

arXiv is making the multiple-identifier problem one identifier worse by creating
arXiv specific identifiers. Deduping articles is a key problem in bibliometrics, 
and we don't want to create a similar deduping problem with author identifiers. 
The OpenID scheme explicitly caters for multiple identifiers for a single person, 
and even for multiple identifiers for each persona a single person might use. 
Facilities to express and leverage multiple identities a described in the 
Yadis/XRDS document~\cite{Yadis1_0}. 

At arXiv we can go some way to addressing this problem by augmenting the 
Atom format machine readable authorship information arXiv exposes with 
correspondences between author identifiers in different schemes. The issue 
then is how to encourage authors to supply and update alternative
identifiers associated with their account. Again we believe that the solution
will be to build useful services, such a links to other systems, that depend
upon this data.

\begin{figure}
\begin{center}
\includegraphics[width=4in]{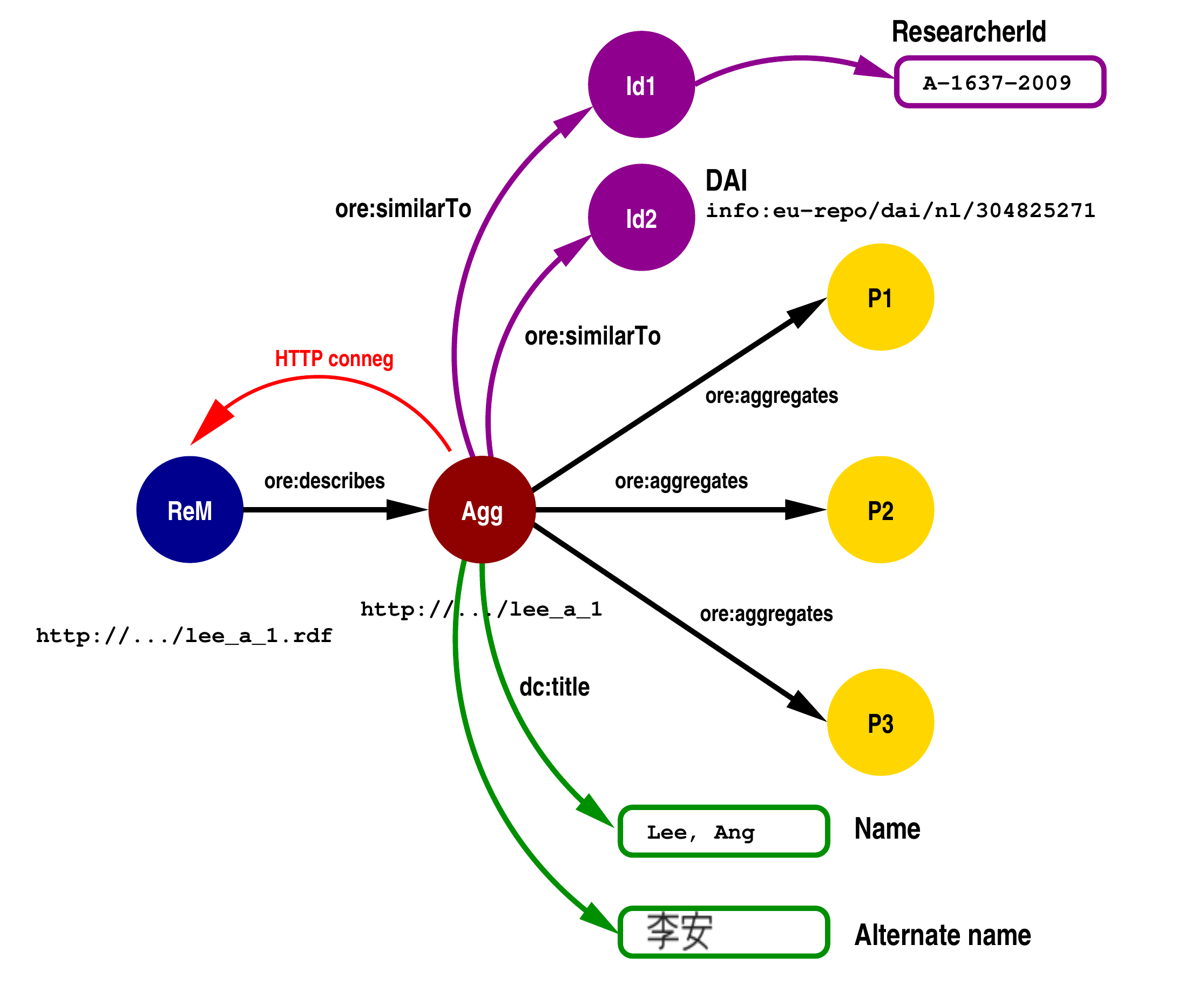}
\caption{\label{fig-ore}Possible OAI-ORE resource map representing the aggregation
of papers by an author Ang Lee. The node ReM is the Resource Map which describes the 
aggregation Agg that has the author id as its URI (\texttt{http://arxiv.org/a/lee\_a\_1}).
The aggregation includes the three papers (P1, P2, P3) authored by this Ang Lee. 
Through the \texttt{ore:similarTo} relation we also indicate two other related 
resources: the identities in other author id schemes. The DAI is a URI and so can 
be related directly as Id2. The ResearcherID is a string and must therefore be 
related via an additional node Id1.}
\end{center}
\end{figure}

Another option to encourage use of arXiv data on the relationships between 
authors, papers authored and identifiers in other schemes is to expose it
in RDF. This is a good application for OAI-ORE Resource Maps~\cite{OAI-ORE} 
and we intend to provide OAI-ORE resource maps as another representation 
available from the arXiv author id. An example showing alternate name, 
article information and how identifiers in other schemes can be exposed 
is illustrated in figure~\ref{fig-ore}.

\section{Conclusions}

There is growing interest in accurate author identification based on explicit
author identifiers. The many different commercial and non-commercial parties
have varying motives and goals, and so are adopting different solutions. It
seems likely that there will continue to be many different systems and multiple
identities for each author. While not perfect, this situation will greatly 
assist the assembly of publication network data linking authors and articles
which will facilitate bibliometric analyses and support new discovery tools
that span multiple repositories. The implementation of author identifiers 
at arXiv, and of services to promote their use has been described to 
illustrate one approach at the repository level. Early use is encouraging
but it remains to be seen how quickly the use of author identification is
adopted and accepted by the scholarly community.

\section{Acknowledgements}

I am pleased to acknowledge contributions from Nathan Woody (Facebook 
and JavaScript interface for arXiv), Thorsten Schwander and Paul Ginsparg. 
This work is supported by Microsoft through a Technical Computing 
Initiative (TCI) Grant. This paper is based on a presentation given
at Open Repositories 2009 on 18 May 2009 but with updated usage data
through September 2009.

\section{Note added in proof}

There has been considerable recent activity around author identifiers, the 
most notable effort being ORCID (Open Researcher Contributor Identification 
Initiative, \url{http://orcid.securesites.net/}) which has significant 
commercial and community participation. Both Elsevier and Thomson Reuters 
are participants in ORCID and the resulting system may replace both Scopus 
Author Id and ResearcherID with a more open and more broadly adopted scheme.

\bibliographystyle{plainurl-sw2009-04-14}
\bibliography{dl}

\end{document}